\documentclass[]{article}
\usepackage{amsmath,amssymb}
\usepackage[spanish,english]{babel}
\begin{document}
\vspace*{-2cm}
\hfill   \hbox{\bf SB/F/02-300}
\hrule \vskip 2cm

\begin{center}
{\Large\bf{$so(2,1)$ scattering via Euclidean Connection for the
Natanzon Hypergeometric Potentials}}\\ \vspace*{0.3cm}
{\large{Hermann Albrecht\footnote{Email: halbrech@fis.usb.ve} 
and Sebasti\'an Salam\'o\footnote{Email: ssalamo@usb.ve} 
\footnote{Accepted for publication in Rev. Mex. F\'{\i}sica, 2002.}}}\\
{\small{Universidad Sim\'on Bol\'{\i}var, Departamento de
F\'{\i}sica,\\ A. P. 89000, Valle de Sartenejas, Miranda 1080-A,
Venezuela.}}
\end{center}

\begin{abstract}
The $so(2,1)$ Euclidean Connection \cite{AlhassidII} formalism is
used to calculate the $S$ matrix for the Hypergeometric Natanzon
Potentials ($HNP$) \cite{NataPNH}.
\end{abstract}

\section{Introduction}

Algebraic scattering has been an active research field for many
years. Pioneer works in the late 1960's and 1970's by Zwanziger
[3] and Barut, et al. [4] dealing with the Coulomb system founded
the basic ideas of this area. More recently, Alhassid et al. [1]
developed an algebraic formalism, known as the Euclidean
connection, to calculate the $S$ matrix for systems whose
Hamiltonian may be written as a function of the Casimir operator
of an $so(2,1)$ algebra.

In this work we study the application of the Euclidean connection
to obtain in a straightforward manner the $S$ matrix for the
$HNP$, by extending the treatment developed in [5].

We start with a  brief review of [5] and then introduce the basic
ideas of the Euclidean connection for $so(2,1)$ [1]. Then we study
its application to the scattering sector of the $HNP$, obtaining
the $S$ matrix for this family of potentials.

\section{Bound states}

As is well known, the Hypergeometric Natanzon Potentials ($HNP$)
[2] are given by:
\begin{eqnarray}\label{eq:PNH}
V(z) &=& \frac{f\,z^2-\left(h_0-h_1+f\right)z+h_0+1}{R(z)}\\
&&+{\left(a+\frac{a+\left(c_1-c_0\right)\left(2 z-1\right)}
{z\left(z-1\right)}-\frac{5}{4}\frac{\Delta}{R(z)}\right)}
\frac{z^2\left(1-z\right)^2 }{R(z)^2}\nonumber
\end{eqnarray}
\noindent{}where
\begin{eqnarray}\label{eq:R(z)DeltaTau}
R(z)=a\,z^2+\tau\, z +c_0&
\Delta =\tau^2-4\,a\,c_0 &
\tau = c_1-c_0-a
\end{eqnarray}
\noindent{}where $f$, $h_{0}$, $h_{1}$, $a$, $c_{0}$, $c_{1}$ are
the {\it{Natanzon Parameters}} and $z(r)$ is a real arbitrary
function that satisfies
\begin{eqnarray}\label{eq:z'(r)}
\frac{d\,z}{dr}\equiv z'=\frac{2\,z\left(z-1\right)}{\sqrt{R(z)}}
&&z(r)\in[0,1]
\end{eqnarray}
\noindent{}and the transformation $r(z)$ is assumed to take
$r\to\infty$ to $z = 1$. In [5] the bound state sector of the
$HNP$ was studied by means of an $so(2,1)$ algebra, obtaining an
algebraic description of it. The realization used is
\begin{eqnarray} \label{eq:RepSO(2,1)}
J_0 &=& -i{\partial_\phi}\nonumber\\ J_{\pm} &=&e^{\pm i\phi}
\left\{\pm\frac{\sqrt{z}(z-1)}{z'} {\partial_ r}
-\frac{1}{2}\frac{i(z+1)}{\sqrt{z}}{\partial_
\phi}\right\}\nonumber\\ && {\pm}e^{\pm
i\phi}\left\{\frac{1}{2}(z-1)\left[\frac{1\mp p_\nu}{\sqrt{z}}-
\frac{z"\sqrt{z}}{z'{}^2}\right]\right\}\nonumber\\
Q&=&\frac{z\left(z-1\right)^2}{z'{}^2}\partial^2_r+
\frac{1}{4}\frac{\left(z-1\right)^2}{4\,z}\partial^2_\phi
+\frac{1}{2}\frac{i\, p_\nu\left(z^2-1\right)}{z}{\partial_\phi}
\nonumber\\
&&+\frac{\left(z-1\right)^2}{4}\frac{\left[z^2\left(2z'''z'-3z"\right)
-z'{}^4\left(p_\nu^2-1\right)\right]}{z\, z'{}^4}
\end{eqnarray}
where $p$ is a parameter, which allows the use of the $so(2,1)$
algebra to completely describe the $HNP$. The algebraic treatment
makes use of
\begin{equation}\label{eq:QSchrodinger}
(Q-q)\Psi(r,\phi)={\cal{G}}(r)(E-H)\Psi (r,\phi)
\end{equation}
\noindent where $H$ is the standard radial Hamiltonian, $q$ is the
eigenvalue of the Casimir operator $Q$ and $\cal{G}$$(r)$ is a
function of $r$. $\Psi (r,\phi)$ is simultaneously an
eigenfunction of $Q$ and $J_0$ as well as of $H$, and is given by
$\Psi (r,\phi)=e^{im\phi}\Phi(r)$. For the bound states the
${\cal{D}}^{(+)}$ representation is used, therefore the compact
generator $J_0$ has eigenvalues [6]
\begin{eqnarray}\label{eq:mPNH}
m=\nu+\frac{1}{2}+\sqrt{q_\nu+\frac{1}{4}}&&\nu=0,1,2,...,\nu_{max}
\end{eqnarray}
\indent{}The energy spectrum is obtained from [5]:
\begin{eqnarray}\label{eq:nuabdPNH}
2\nu +1 &=& \alpha_\nu -\beta _\nu - \delta _\nu
\end{eqnarray}
\noindent{}where
\begin{eqnarray}\label{eq:abdPNH}
\alpha _\nu &\equiv& \sqrt{-aE_\nu+f+1} =p_\nu+m_\nu \nonumber\\
\beta_\nu&\equiv&\sqrt{-c_0E_\nu+h_0+1}=p_\nu-m_\nu \nonumber\\
\delta_\nu&\equiv&\sqrt{-c_1E_\nu+h_1+1}=\sqrt{4q_\nu+1}\nonumber
\end{eqnarray}
This are the basic equations of the algebrization derived in [5].

\section{Euclidean connection}
The main motivation for Euclidean connection [1] is that the
asymptotic states (regular solution) are written as [7]:
\begin{equation}\label{eq:AsymptState}
|\phi\rangle^{(\infty)} = {\cal{A}}e^{ikr} + {\cal{B}}e^{-ikr}
\end{equation}
\noindent{}where ${\cal{A}}$ and ${\cal{B}}$ are the Jost
functions and $e^{\pm{}ikr}$ are eigenfunctions of an Euclidean
algebra $e(2)$:
\begin{eqnarray}\label{eq:ConmuE(2)}%
\left[L_z,P_x\right] = iP_y & \left[L_z, P_y \right]=-iP_x &
\left[ P_x, P_y\right] = 0
\end{eqnarray}
A realization in polar coordinates of the ladder and Casimir
operators is as follows:
\begin{eqnarray}\label{eq:PpmCasPolares}
P_\pm &=&e^{\pm i\phi}\left[-i\partial_{r}+\frac{1}{{r}}
\left(\pm\,i\partial_\phi+\frac{1}{2} \right)\right]\nonumber\\
{\bf{P}}^2&=& -\partial_{r}^2-\frac{1}{{r}^2}
\left(\partial_\phi^2+\frac{1}{4}\right)
\end{eqnarray}
\noindent{}where $P_\pm\equiv{}P_x\pm{i}P_y$ and
${\bf{P}}^2\equiv{}P_{+}P_{-}$. Taking the limit $r\to\infty$, the
asymptotic generators are obtained:
\begin{eqnarray}\label{eq:RepPolAsintE(2)}
P_\pm^{(\infty)} = -ie^{\pm i\phi}\partial_{r} && L_z^{(\infty)} =
-i\partial_\phi\nonumber\\
{\bf{P}}^{2\;(\infty)}=-\partial_{r}^2
\end{eqnarray}
\noindent{}that also closes an $e(2)$ algebra. The eigenstate
basis $\left\{|\pm k,m\rangle^{(\infty)}\right\}$ is
\begin{equation}\label{eq:e(2)basis}
|\pm k,m\rangle^{(\infty)} = e^{\pm ik{r}}e^{im\phi}
\end{equation}
\noindent{}and the action of the asymptotic generators upon it is
given by:
\begin{eqnarray}\label{eq:pmkmE(2)}
P_\pm^{(\infty)} |(\pm)k,m\rangle^{(\infty)} &=& (\pm)k|(\pm)k,
m\pm 1\rangle^{(\infty)}\nonumber\\
{\bf{P}}^2|\pm{k},m\rangle^{(\infty)} &=&
k^2\,|\pm{k},m\rangle^{(\infty)}
\end{eqnarray}
As is well known, we can expand the ladder operators of $so(2,1)$
in terms of the $e(2)$ generators [8]. Therefore, given an
$so(2,1)$ algebrization of the bound states, it is natural to
study its extension to the scattering sector by means of the
mentioned expansion. The expanded ladder operator
$J_{+}^{(\infty)}$ of $so(2,1)^{(\infty)}$ in the continuous
representation (where $j=-\frac{1}{2}+i\,f(k)$) obtained in [1]
is:
\begin{equation} \label{eq:J+inftyE(2)}
J_{+}^{(\infty)}(\pm k) = \frac{e^{i\gamma_\pm(k)}}{\pm k}
\left[\left(-\frac{1}{2} \mp if(k)\right)P_{+}^{(\infty)}
+L_z^{(\infty)}P_{+}^{(\infty)}\right]
\end{equation}
where $f(k)$ is defined by:
\begin{eqnarray}\label{eq:k=h(f(k))}
k^2 = h\left(-f^2(k)\right)&& E_j = h\left[q+\frac{1}{4}\right]
\end{eqnarray}
\noindent{}where $h(\eta)$ is the function that connects the
Energy $E_j$ with the Casimir eigenvalue $q$ or, equivalently, the
Hamiltonian with the Casimir operator $Q$.

A recurrence relation may be calculated and solved for the
coefficients $A_m$ and $B_m$ by applying (\ref{eq:J+inftyE(2)})
directly to (\ref{eq:AsymptState}), with $|\phi\rangle^{(\infty)}=
e^{-im\phi}|j,m\rangle^{(\infty)}$, and equating it to the action
of $J_{+}^{(\infty)}$ upon $|j,m\rangle^{(\infty)}$. Since
$S_m(k)= A_m/B_m$, then:
\begin{equation}\label{eq:Rm(k)}
S_m(k) = e^{im\left[\gamma_{+}+\gamma_{-}\right]}
\frac{\Gamma\left[m+\frac{1}{2}-if(k)\right]}
{\Gamma\left[m+\frac{1}{2}+if(k)\right]}\Delta(k)
\end{equation}
\noindent where $\Delta(k)$ is a constant factor to be determined.

\section{$S$ matrix for the $HNP$}

Starting from (\ref{eq:RepSO(2,1)}), the asymptotic algebra is
readily obtained by taking the limit $r\to\infty$ (or equivalently
$z=1$).
\begin{eqnarray}\label{eq:RepSO(2,1)Asint}
J_0^{(\infty)} &=& -i\,{\partial_\phi}\\ J_{\pm}^{(\infty)} &=&
e^{\pm i\phi} \left[\mp\frac{\sqrt{c_1}}{2}\, {\partial_ r}
-i\,{\partial_\phi}\,\pm\frac{1}{2}\right]\nonumber\\ Q^{(\infty)}
&=& \frac{c_1}{4}\partial_r^2-\frac{1}{4}\nonumber
\end{eqnarray}
\indent{}Since these generators form also an $so(2,1)$ algebra,
the Euclidean connection may be applied directly. Equation
(\ref{eq:RepSO(2,1)Asint}) corresponds to the $so(2,1)^{(\infty)}$
algebra mentioned before. We only need to determine the expansion
coefficients $f(k)$ and $\gamma_\pm(k)$. This is accomplished by
equating the action of (\ref{eq:J+inftyE(2)}) with the one of
(\ref{eq:RepSO(2,1)Asint}), obtaining
\begin{eqnarray}\label{eq:ParExp}
\gamma_\pm(k) = 0 && f(k) = \frac{k\sqrt{c_1}}{2}\\
&\Rightarrow& h'(\eta) = -\frac{4}{c_1}\;\eta\nonumber
\end{eqnarray}
From (\ref{eq:nuabdPNH}) we know that
\begin{eqnarray}\label{eq:EjPNH}
E_j = -\frac{4\,q -h_1}{c_1}
&=&h\left[q+\frac{1}{4}\right]\nonumber\\
&\Rightarrow& h(\eta) \equiv-\frac{4}{c_1}\;\eta
+\frac{h_1+1}{c_1}
\end{eqnarray}
Since we are dealing with the scattering sector and
$V_{PNH}\propto{h_1+1}$ as $r\to\infty$ [2], $h_1=-1$. Therefore
$h'(\eta)=h(\eta)$ and the $S$ matrix is given by
(\ref{eq:Rm(k)}):
\begin{equation}\label{eq:RmPNHE(2)}
S_m(k)={\frac{
\Gamma\left(m+\frac{1}{2}-\frac{1}{2}\,ik\sqrt{{c_1}}\right)}
{\Gamma\left(m+\frac{1}{2}+\frac{1}{2}\,ik\sqrt{c_1}\right) }}
\Delta_{m}(k)
\end{equation}
\noindent{} where $\Delta_{m}(k)$ is a regular function of $k$
derived by Natanzon [2]. It can easily be verified that the poles
are in agreement with the bound state energy spectrum.

\section{Conclusions}

We have successfully extended the algebraic treatment for the
bound states developed in [5] to the scattering sector via the
$so(2,1)$ Euclidean connection, therefore obtaining a purely
algebraic description of the $HNP$ using an $so(2,1)$ algebra.
This can also be accomplished with a methodology analogous to the
one used by Frank, et al. [9]. This was done in detail and it is
to be published soon [10]. Meanwhile, a short review may be found
in this volume [11]. Even though Alhassid, et al. had obtained an
algebraic description of the $HNP$ in [12] by means of an
$so(2,2)$ algebra, it is important to notice that our treatment is
simpler.

\vspace*{1cm}
\noindent{\large\bf{Acknowledgments}}

This work has been partially supported by grants USB-61D30 
and FONACIT No. 6-2001000712.

\end{document}